# Photoluminescence of ion-synthesized 9R-Si inclusions in SiO$_2$/Si structure: effect of irradiation dose and oxide film thickness

Alena Nikolskaya, Alexey Belov, Alexey Mikhaylov, Anton Konakov, David Tetelbaum, Dmitry Korolev [a]

*Research Institute of Physics and Technology, Lobachevsky University, Nizhny Novgorod 603022, Russia*

The study of hexagonal silicon polytypes attracts special attention due to their unique physical properties compared to the traditional cubic phase of Si. Thus, for some hexagonal phases, a significant improvement in the emission properties has been demonstrated. In this work, the luminescent properties of SiO$_2$/Si structures irradiated with Kr$^+$ ions at different doses and annealed at 800 °C have been systematically investigated. For such structures, a photoluminescence line at ~ 1240 nm is observed and associated with the formation of hexagonal 9R-Si phase inclusions. It is found that the variation in the thickness of oxide film and the relative position of ion distribution profile and film / substrate interface leads to a regular change in the luminescence intensity. The nature of the observed dependencies is discussed as related to the interplay between the desirable 3C-Si → 9R-Si structural transition and generation of nonradiative defects. The revealed regularities suggest optimal ion irradiation conditions for synthesis of optically active 9R-Si phase in diamond-like silicon.

---

[a] Electronic mail: dmkorolev@phys.unn.ru

For many decades, silicon has been and remains the main material for micro- and nanoelectronics. Its wide reserves, the available technology for producing perfect crystalline wafers of large diameter, and unique physical properties make it promising for the use in new generation semiconductor devices. However, the applicability of silicon in optoelectronic and integrated optics devices is limited due to the fundamental lack of this semiconductor – the indirect gap of its energy structure. A promising approach for improving the silicon emission properties can be the use of other polytypes, in particular hexagonal ones, for which the possibility to "straighten" the energy structure relative to that of cubic silicon was theoretically shown [1].

Among the most well-known methods for the synthesis of hexagonal phase, the methods of deposition of thin silicon layers on various substrates can be distinguished [2-5], for which the formation of new phase inclusions is caused by stresses, the sources of which are the difference in the crystal lattice parameters of the film and the substrate. Another well-known method is the growth of silicon nanowires, the formation of a hexagonal phase in which is associated with the strong surface stresses arising during the growth [6-10]. However, it was noticed [7, 11] that the synthesized polytypes are often unstable and eventually relax into a cubic structure.

Along with a change in the crystal structure, the strong luminescence was found for the samples containing inclusions of hexagonal silicon in the visible and near-IR spectral regions with the intensity noticeably exceeding that for the luminescence of diamond-like silicon [9, 10]. However, the complexity of methods used for the synthesis of hexagonal silicon phases and their instability limit the practical application of such structures.

We have previously demonstrated that, by ion implantation into $SiO_2$/Si structures with subsequent annealing, it is possible to form the inclusions of hexagonal 9R-Si phase in Si substrate at the interface with $SiO_2$ film, which reveal luminescence at a wavelength of ~ 1235 nm [12-14]. In the present work, the influence of ion implantation parameters on the photoluminescence spectra is systematically studied in dependence on mutual arrangement of the implanted layer and the film / substrate interface.

The initial samples were *n*-Si (100) wafers with a resistivity of 4.5 Ω×cm. $SiO_2$ films with thicknesses of 50-220 nm were grown by thermal oxidation in dry oxygen. The $SiO_2$/Si samples were irradiated with ions of an inert gas $Kr^+$ with an energy of 80 keV and doses of $1\times10^{16}$, $5\times10^{16}$, and $1\times10^{17}$ $cm^{-2}$. The experimental conditions used provided a different arrangement of the distribution profiles of ions and radiation defects, calculated using the SRIM program [15], relative to the film / substrate interface (Figure 1). Post-implantation annealing was carried out at 800 °C in a dried nitrogen atmosphere (30 min). The photoluminescence (PL) spectra were studied at 77 K with excitation by a laser with a wavelength of 408 nm.

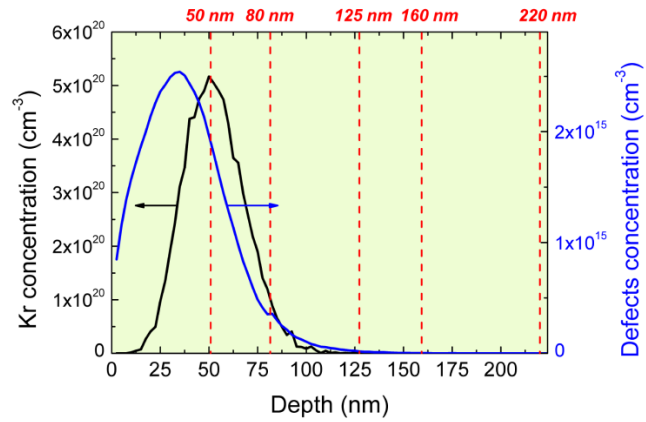

FIG. 1. Distribution profile of implanted $Kr^+$ atoms (80 keV, dose $5\times10^{16}$ cm$^{-2}$) and radiation defects. Dashed lines mark the film / substrate interfaces for the $SiO_2$ film thicknesses used in the experiment.

In our previous work, the behavior of PL line associated with the appearance of 9R-Si phase inclusions was investigated for certain $SiO_2$ film thicknesses [13]. It was shown that the appearance of a PL line at ~ 1240 nm correlated with the presence of this phase in the cross-section transmission electron microscopy images. However, the cases of small film thicknesses were not investigated, when the interstitial Kr atoms and recoil atoms from $SiO_2$ films could penetrate in sufficient quantities into the Si substrate.

Let's first consider the regularities of PL properties of irradiated $SiO_2$/Si structures depending on the thickness of the oxide film. The PL spectra of $SiO_2$/Si samples with different thicknesses of $SiO_2$ films irradiated with $Kr^+$ and annealed at 800 °C are shown in Figure 2 for three different irradiation doses. For the lowest dose used ($1\times10^{16}$ cm$^{-2}$), a weak luminescence is observed for the sample with the minimum film thickness (50 nm). With the increase in thickness, the PL intensity increases and reaches its maximum for a sample with a 125 nm film thickness, then decreases and completely disappears for a sample with the maximum film thickness. When $Kr^+$ is implanted with a dose of $5\times10^{16}$ cm$^{-2}$, the PL is detected only starting from 80 nm film thickness, reaches maximum for a sample with 160 nm film thickness, and is present as a weak PL band for a sample with 220 nm $SiO_2$ thickness. Finally, for the highest dose used, luminescence is observed only for the samples with 125 and 160 nm film thicknesses and is practically not observed for a sample with 220 nm $SiO_2$ thickness. Luminescence in the region corresponding to the interband transitions in 3C-Si (~ 1.1 μm) is not observed for all the samples under study.

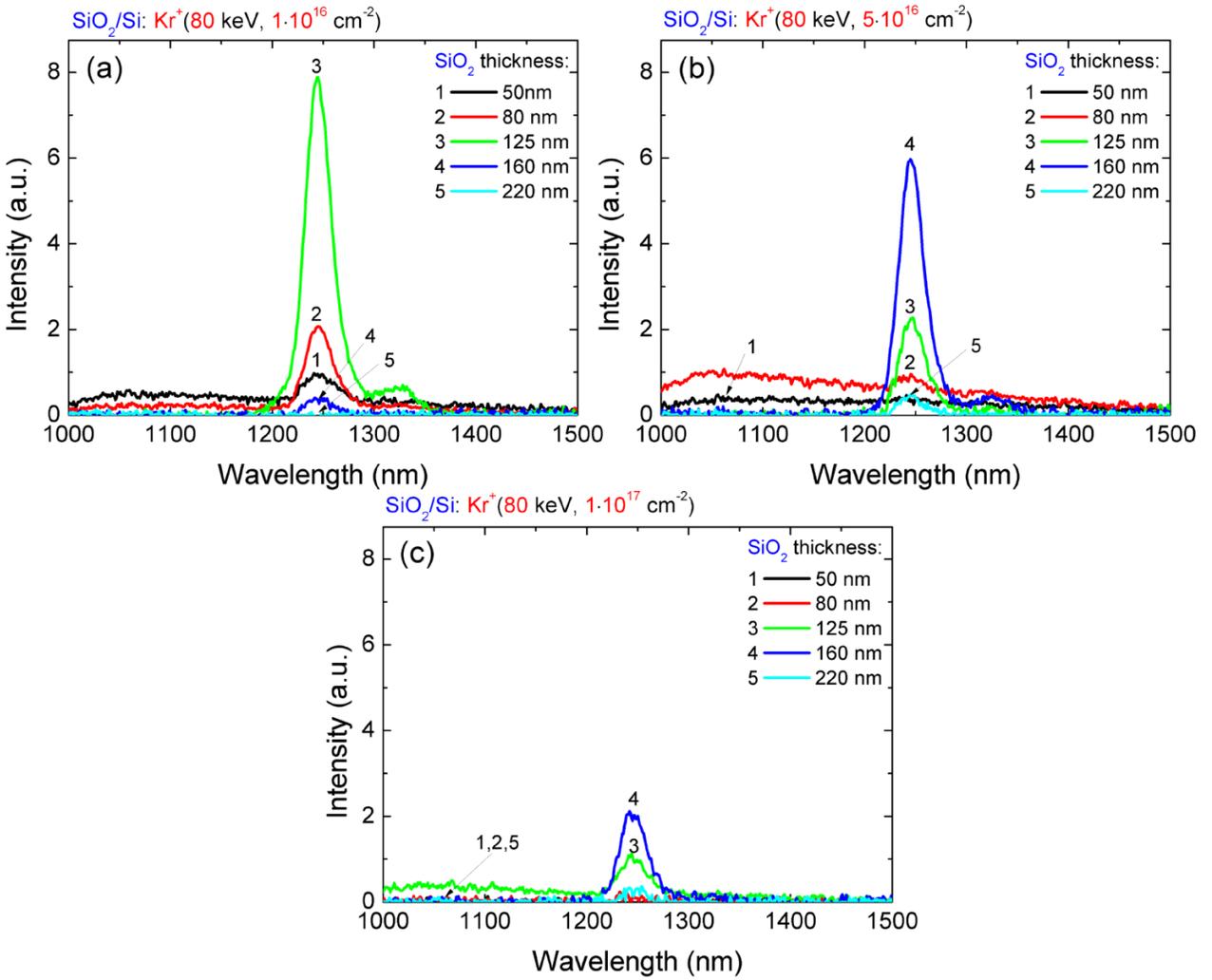

FIG. 2. PL spectra (77 K) of SiO$_2$/Si samples with different SiO$_2$ films thicknesses irradiated with Kr$^+$ ions with an energy of 80 keV and various doses of 1×10$^{16}$ (a), 5×10$^{16}$ (b), and 1×10$^{17}$ cm$^{-2}$ (c) after annealing at 800 °C (30 min).

The change in PL intensity vs. the dose of Kr$^+$ ions differs significantly for the SiO$_2$/Si samples with different SiO$_2$ film thicknesses (Figure 3). For small film thicknesses (50, 80, and 125 nm), there is a tendency for the PL intensity to decrease with an increase in the ion dose. For a sample with 160 nm film thickness, a significant increase in the PL intensity is observed with an increase in the Kr$^+$ dose to 5×10$^{16}$ cm$^{-2}$, followed by a decrease for the dose of 1×10$^{17}$ cm$^{-2}$. For a sample with 220 nm film thickness, the same trend persists; however, the PL intensity turns out to be very low for all the irradiation doses used.

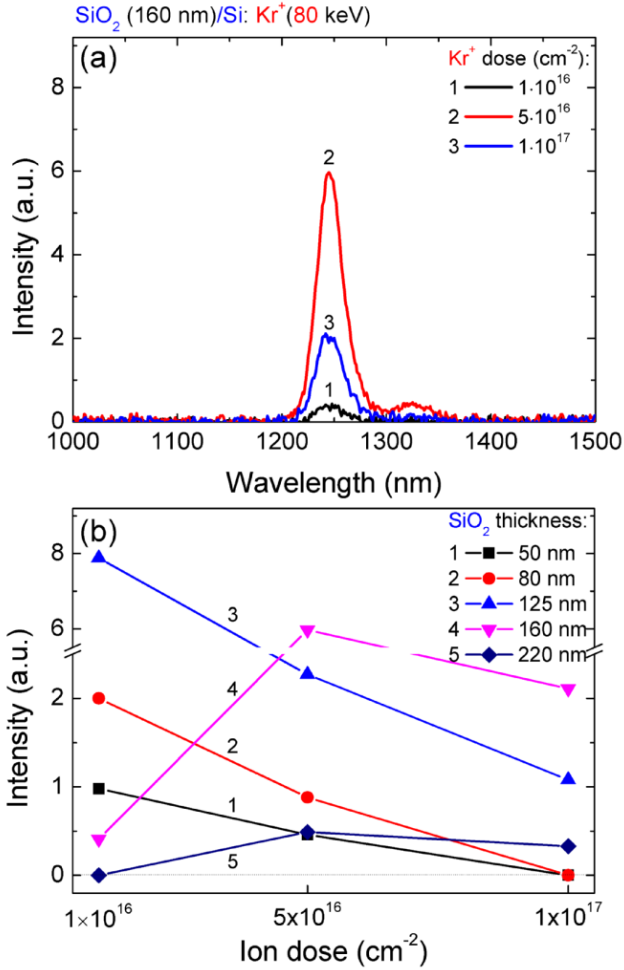

FIG. 3. PL spectra of $SiO_2$ (160 nm)/Si samples for different doses of $Kr^+$ ions (a) and dependences of the PL intensity maxima at ~ 1240 nm on the ion dose for irradiated $SiO_2$/Si samples with different oxide film thicknesses (b). Solid lines in Fig. 3b are the guides of the eye.

Let's consider the nature of the observed regularities of changes in the PL intensity with variations in the $SiO_2$ film thickness and the dose of implanted $Kr^+$ ions. The used experimental conditions were chosen to vary two factors: the relative position of the implanted layer and the film / substrate interface, as well as the number of atoms implanted into the sample. In reality, these factors are interdependent, and it is difficult to differentiate their contributions. In the case of sufficiently small film thicknesses (50 and 80 nm), it can be noted from Figure 1 that, for the used irradiation parameters, both the distribution profiles of implanted ions and the distribution of defects cross the interface between the $SiO_2$ film and the Si substrate. For a 125 nm film thickness, the overlap of the distribution profiles of both ions and defects with the interface is much less pronounced, but still present. For these samples, the PL spectra show a decrease in intensity with an increase in the dose of $Kr^+$ ions. This may be explained as follows. As we suggested earlier [13], the formation of the 9R-Si phase responsible for the luminescence at ~ 1240 nm occurs due to the relaxation upon annealing of mechanical stresses arising during ion irradiation.

This fact is supported by the 3C-Si → 9R-Si structural transition in the near-surface silicon layer of the substrate observed in the cross-sectional transmission electron microscopy images. At the same time, the O and Si recoil atoms, which penetrate from the film into the substrate, take part in this process [16]. This, on the one hand, promotes the formation of the 9R-Si phase and the appearance of the corresponding luminescence line at ~ 1240 nm, and, on the other hand, an increase in the ion dose leads to a significant increase in the number of defects in Si, which quench the PL due to nonradiative recombination. This effect is especially pronounced for a sample with 125 nm oxide film thickness. As can be seen from Figure 1, the defect profile is shifted relative to the distribution profile of implanted atoms towards shallower depths. This leads to a decrease in the number of defects in the region where the PL centers are located, and the quenching of luminescence is less pronounced. With an increase in the dose of implanted ions, the number of ions that reach the interface between the film and the substrate and create recoil atoms in Si increases; the concentration of defects increases, which leads to a gradual quenching of the luminescence.

For the $Kr^+$-irradiated $SiO_2$/Si samples with 160 and 220 nm oxide film thicknesses, the situation is different. In this case, the distribution profiles of implanted atoms and defects are located far from the film / substrate interface; therefore, both implanted atoms and defects in Si (caused by Si and O recoil atoms) have practically no direct effect on the Si substrate. The regularities of the change in PL intensity at ~ 1240 nm in this case differ from those for the case of thinner films. For a film with 160 nm thickness, a nonmonotonic dependence of the PL intensity on the dose of $Kr^+$ ions is observed, with maximum at a dose of $5\times10^{16}$ cm$^{-2}$. This nonmonotonous behaviour is associated with the competing influence of two factors on the PL intensity: on the one hand, with increasing dose, the mechanical stresses increase, under the influence of which a phase transition to the 9R-Si polytype occurs, and, on the other hand, the concentration of defects in the substrate, which are the nonradiative recombination centers, increases. At smaller thicknesses, the second factor predominates even at low doses, therefore, for them, the intensity monotonically increases with the dose, while it begins to affect only at a sufficiently high dose for the 160 nm thickness. If the film is too thick (220 nm), only a small number of ions reaches the interface that the first factor becomes dominant for all studied doses, although the degree of phase transition in the substrate is small due to the remoteness of the interface from the irradiated layer – relaxation of elastic stresses occurs mainly within the film.

To summarize, the effect of the film thickness in the $SiO_2$/Si structure, as well as the irradiation dose of $Kr^+$ ions on the photoluminescence at a wavelength of ~ 1240 nm, which is associated with the formation of the 9R-Si phase in the silicon substrate at the interface with the $SiO_2$ film, has been studied. It is shown that the luminescent properties of the irradiated samples are determined by the mutual arrangement of the distribution profiles of implanted ions and radiation defects, and the interface between the $SiO_2$ film and Si substrate. The main mechanism for the synthesis of luminescence centers is the formation of the 9R-Si phase in silicon substrate upon annealing as a result of relaxation of mechanical stresses arising in the $SiO_2$/Si structure during irradiation with $Kr^+$ ions. At the same time, there is an optimal range of thicknesses and doses for which this effect is most pronounced.

This study was supported by the Lobachevsky University competitiveness program in the frame of 5-100 Russian Academic Excellence Project.

**DATA AVAILABILITY**

The data that support the findings of this study are available from the corresponding author upon reasonable request.

and D.I. Tetelbaum, Surf. Coatings Technol. **386**, 125496 (2020).

[15] See http:// www.srim.org for more information about SRIM (The Stopping and Range of Ions in Matter).

[16] A.A. Nikolskaya, D.S. Korolev, A.A. Konakov, A.N. Mikhaylov, A.I. Belov, M.O. Marychev, R.I. Murtazin, D.A. Pavlov, and D.I. Tetelbaum, J. Phys. Conf. Ser. **1695**, 012031 (2020).